\documentclass[aip,preprint, english,showpacs,superscriptaddress]{revtex4}
\usepackage{bm}
\usepackage{epsfig}
\usepackage{color}
\usepackage[english]{babel}

\makeatother

\begin{document}
\newcount\timehh \newcount\timemm \timehh=\time \divide\timehh
by 60 \timemm=\time \count255=\timehh\multiply\count255 by -60
\advance\timemm by \count255

\title{ 
Room-temperature transport of indirect excitons  in  (Al,Ga)N/GaN quantum wells. }

\author{F. Fedichkin}
\affiliation{Laboratoire Charles Coulomb, UMR 5221
 CNRS-Universit\'{e} de Montpellier, 34095 Montpellier, France}
 \author{T. Guillet}
\affiliation{Laboratoire Charles Coulomb, UMR 5221
 CNRS-Universit\'{e} de Montpellier, 34095 Montpellier, France}
 \author{P. Valvin}
\affiliation{Laboratoire Charles Coulomb, UMR 5221
 CNRS-Universit\'{e} de Montpellier, 34095 Montpellier, France}
\author{B. Jouault}
\affiliation{Laboratoire Charles Coulomb, UMR 5221
 CNRS-Universit\'{e} de Montpellier, 34095 Montpellier, France}
 \author{C. Brimont}
\affiliation{Laboratoire Charles Coulomb, UMR 5221
 CNRS-Universit\'{e} de Montpellier, 34095 Montpellier, France}
 \author{T. Bretagnon}
\affiliation{Laboratoire Charles Coulomb, UMR 5221
 CNRS-Universit\'{e} de Montpellier, 34095 Montpellier, France}
 \author{L. Lahourcade}
\affiliation{Institute of Physics, EPFL, CH-1015 Lausanne, Switzerland}
\author{N. Grandjean}
\affiliation{Institute of Physics, EPFL, CH-1015 Lausanne, Switzerland}
\author{P. Lefebvre}
\affiliation{Laboratoire Charles Coulomb, UMR 5221
 CNRS-Universit\'{e} de Montpellier, 34095 Montpellier, France}
 \author{M. Vladimirova}
\affiliation{Laboratoire Charles Coulomb, UMR 5221
 CNRS-Universit\'{e} de Montpellier, 34095 Montpellier, France}
 %


\begin{abstract}

We report on the exciton propagation in polar  (Al,Ga)N/GaN quantum wells over several micrometers and up to room temperature.
The key ingredient to achieve this result is the crystalline quality of GaN quantum wells (QWs) grown
on GaN template substrate.
By comparing micro-photoluminescence images of two identical QWs  grown on  sapphire and on GaN, we 
reveal the twofold role played by GaN substrate in the  transport of excitons.
First, the lower threading dislocation densities in such structures yield higher exciton  radiative efficiency, thus limiting non-radiative  losses of propagating excitons.
Second, the absence of the dielectric mismatch between the substrate and the epilayer strongly limits the photon guiding effect in the plane of the structure,
making exciton transport easier to distinguish from photon propagation.
Our results pave the way towards room-temperature gate-controlled  exciton transport in wide-bandgap polar heterostructures.
\end{abstract}

\maketitle
\emph{Introduction.} 
An indirect exciton (IX) in a semiconductor quantum-well (QW)  is a quasiparticle composed of an electron and a hole separated  along the growth direction, but still bound by Coulomb interaction. IXs can be realised in both wide single QWs  \cite{Polland} and in coupled QWs \cite{Armiento}. 
The prominent feature of IXs is their long lifetime,  exceeding lifetimes of regular direct excitons by orders of magnitude  \cite{Fukuzawa,ButovShaskin}. Their long lifetimes allow IXs to travel over large distances before recombination, providing the opportunity to study exciton transport by optical imaging \cite{Voros2005PRL,Remeika2009,Leonard2012}. Moreover, they possess strong permanent dipole moments along the growth axis, which 
 is essential for electrical control of excitonic fluxes, and for excitonic 
 device operation \cite{Grosso2009,Leonard2012,High2007OptLett,HighOptLett2010,Andreakou2014APL}.
 IXs were most extensively studied in GaAs-based QWs  \cite{Armiento,Butov1998,Voros2005PRL,Cohen2011PRL,Alloing,Gartner,Violante2014}, but recently their realisation in 
 wide bandgap semiconductors such as GaN \cite{Fedichkin2015}, and ZnO \cite{Kuznetsova2015}, has been considered.
GaN/(Al,Ga)N and ZnO/(Zn,Mg)O wide  QWs grown along the (0001) crystal axis, naturally exhibit built-in electric fields up to MV/cm \cite{Leroux1998PRB,Esmaeili}, so that IXs are naturally created in the absence of any external electric field.
Such IXs have binding energy of tens of meV, largely above the 4 meV reached in a typical GaAs-based structure \cite{Morhain2005PRB,Bigenwald}. 
This may extend the operation of the excitonic devices up to room temperature. 
Furthermore, such materials are promising candidates for the realisation of  coherent states of IXs at temperatures higher than a fraction of Kelvin \cite{HighNature,Alloing2013}. These properties make materials with robust IXs particularly interesting. 
Nevertheless, so far, even  in these materials with robust excitons there were no evidence of room-temperature transport \cite{Fedichkin2015,Kuznetsova2015}.

In this Letter we demonstrate excitonic transport  up to room temperature over distances up to 
$5-10$~$\mu$m in GaN/(Al,Ga)N  QW grown on GaN substrate. 
By comparing time-resolved photoluminescence (PL) and spatially resolved micro-photoluminescence ($\mu$PL) images of two identical QWs grown, respectively,  on GaN and on sapphire, we 
reveal the important role played by the substrate in exciton propagation.
First, the radiative efficiency appears to be higher for GaN-grown structures, with much weaker temperature-induced quenching, presumably due to lower threading dislocation densities. 
Second, the absence of dielectric mismatch between the substrate and the epilayer strongly limits the photon guiding effect along the plane of the structure,
making exciton transport easier to distinguish from photon propagation  \cite{Fedichkin2015}.
Careful benchmarking of spatial and temporal dynamics suggests that even in the best quality 
QWs grown on GaN substrate, most of exciton 
propagation in the optically excited  QW takes place in the  regime where built-in electric field is efficiently screened by the presence of relatively high densities of dipolar IXs.
These results constitute an important building block for our understanding of exciton transport in wide-bandgap polar heterostructures, and pave  the way towards room-temperature gate-controlled  exciton transport.

\emph{Samples and experimental setup.} 
The studied structures are $7$~nm-wide  single GaN QWs grown along the 
($0001$) axis. QWs are embedded in 
45~nm~/~140~nm Al$_{0.19}$Ga$_{0.81}$N barriers, grown on either sapphire (Sample 1) or  GaN (Sample 2) 
substrate  following the deposition of $3$~$\mu$m thick templates. 
The schematic band diagram of the structure is given in Fig.~1(a). 
Due to the difference in spontaneous and piezoelectric polarization between the well and barrier materials, interfacial fixed charges accumulate, resulting in a a built-in electric field estimated as $1$~MV/cm in this structure \cite{Lefebvre2004}.
The built-in electric field pushes the electron and the hole toward opposite interfaces of the QW, breaking the symmetry of the excitonic wavefunction, and inducing a permanent dipole moment along the growth axis. 
Optical excitation of 
the structure results in the screening of the built-in field,  and in a blue shift $E_{BS}$ of the exciton energy  $E_X$ with respect to $E_0$, the energy of the exciton ground state in the absence of photo-excitation (Fig.~1(b)).
The dependence of $E_{BS}$ on the exciton density $n$ is estimated by solving numerically the coupled Schr\"odinger-Poisson equations by the
self-consistent procedure described in Ref.~\onlinecite{Lefebvre2004}.
The result of this calculation is nicely fitted by a linear law: 
$E_{BS}=\Phi_0 n$, where $\Phi_0=10^{-13}$~eV/cm$^2$. 
The energy of the unscreened QW $E_0 \approx 3.0$~eV was obtained numerically and measured via time-resolved experiments presented below.

In $\mu$PL experiments, a cw laser beam at $\lambda=266$~nm was focused onto a $1$~$\mu$m spot on the sample surface. 
In time-resolved measurements, excitons were photogenerated by frequency-tripled Ti-sapphire laser at the same wavelength (pulse duration $150$~fs, repetition rate $8$~kHz, average power $660$~nW) and using a  $\sim 100$~$\mu$m excitation spot diameter. 
Spectrally-resolved images were obtained using a spectrometer equipped with a 150~gr/mm grating blazed at 390~nm. The detector for spatially resolved images was a  CCD camera with $1024 \times 512$ pixels and pixel size  of 13~$\mu$m, which corresponds to $0.405$~$\mu$m on the sample surface. Time-resolved measurements of the emission kinetics were performed using a Hamamatsu streak camera (model C10910 equipped with an S20 photocathode and UV coupling optics), enhanced for ultraviolet detection.

\emph{Experimental results and discussion.} 
Fig. 2 shows  $\mu$PL spectra  (colour-encoded in log scale) as a function of the distance $r$ from the excitation spot. 
Images were taken for the two samples and at three different temperatures
(excitation power $P=1.3$~mW).
Excitonic recombination in the GaN template and/or substrate, near $E=3.48$~eV  is rapidly quenched by increasing temperature. 
Below this energy the emission is dominated by QW excitons, with
several phonon replicas below the zero-phonon line.
At 30 K in both samples one can clearly see the arrow-shaped pattern, characteristic of exciton propagation, accompanied by the  decrease of the emission intensity and energy \cite{Fedichkin2015,Voros2005PRL}. 
The spatial extension of the arrow-shaped pattern is similar for the two samples.
While in Sample 1 this pattern is completely washed out at $300$~K,  in Sample 2 exciton emission and propagation persist up to room temperature. 
%
%
%
Two other prominent features can be observed in Sample 1 only. 
These are (1) interference fringes and (2) a strong emission tail with spatially independent energy at long distances (above 30 $\mu$m).
Both features have already been observed in samples grown on sapphire substrates \cite{Fedichkin2015}.
We attributed the fringes to Fabry-Perot interferences due to the reflection at the GaN/sapphire interface, and the emission tails  to the emission of excitons created by a secondary absorption process.
Indeed, a significant part of the higher-energy photons emitted around $r = 0$ by the QW itself is guided along the sample due to high refractive index contrast between GaN and sapphire, as well as between (Al,Ga)N and air. These photons are capable of creating, at larger distances, secondary excitons (Fig. 1(c)). 
In Sample 2 interference fringes  are suppressed, because it does not have GaN/Sapphire interface.
Nevertheless, even for Sample 2, a non-zero density of secondary photons is still present along the sample plane.  
This can be clearly seen in Fig. 3(a, b), where colour-encoded $\mu$PL
intensity spatial maps normalized by power  are shown at $10$~K for two different excitation powers: the tails observed at $3.4$~eV (a) and $3.2$~eV (b) are attributed to scattered secondary photons produced by the exciton emission at $r=0$.
The colour scale is the same as in Fig.~2, and the acquisition time is $10$ times longer. 
We argue that propagation of secondary photons along the sample plane remains possible, although less efficient than in Sample 1, probably due to total internal reflection at the sample surface (Fig. 1(d)). 
The fluctuating energy tail is particularly visible in Fig.~3(b) at $3.05$--$3.1$~eV, with a clear jump of emission energy at $r\approx -30$~$\mu$m.  Like in  Sample 1, this low-intensity emission is related to the 
secondary excitons created far from the  laser excitation spot.  
These excitons contribute to the screening of the built-in electric field and of the in-plane potential fluctuations resulting from the disorder, and thus are important for the transport of excitons  created directly under the laser spot  \cite{Ivanov:2002,Remeika2009,Remeika2015}.
%
%
%
%

Because photon guiding is only partly suppressed in the absence of GaN/sapphire interface, it is important to evaluate the energy of the unscreened structure in an independent
time-resolved experiment. 
For this purpose, we used a pulsed laser excitation with low repetition frequency and low average power. 
%
Typical time-resolved PL spectra  measured in Sample 2 at $T=10$~K and at different time delays with respect to the excitation pulse are shown in Fig.~4~(a).
One can see that the QW emission energy $E_X$ goes from $3.06$~eV down to $E_0\approx 3.0$~eV,
the value that should thus be considered as the energy of the unscreened structure.
The integrated intensity of exciton emission as a function of time, and the blue-shift of the zero-phonon PL line with respect to $E_0$ are shown by symbols in Figs. 4~(b) and 4~(c), respectively. 
The blue-shift exponentially decreases  towards $E_0$ with a time constant of $2$~$\mu$s, as shown by a solid line in Fig.~4(c). 
%
%
After the same time transient the decay of the integrated intensity reaches an exponential behavior with 
$\tau(n=0)=10$~$\mu$s, shown by a solid line in  Fig.~4~(b).
This slow decay of integrated intensity corresponding to almost constant emission energy $E_0$ provides the estimation of exciton radiative lifetime in the unscreened QW with zero exciton density. 
Similar analysis in Sample 1 gives $\tau(n=0)=22$~$\mu$s and  $E_0\approx 3.08$~eV. 
The reasons for this difference may be related to slight 
non intentional variations in the QW width and in the barrier composition.
From this analysis we deduce that all the emission observed in cw experiments presented here
 stems from QW regions were original built-in electric field and its fluctuations are partly 
 screened due to photoexcitation. A significant density of excitons exceeding $10^{11}$~cm$^{-2}$ is therefore present along the QW plane  up to $100$~$\mu$m  away from the excitation spot.
 %


On the basis of this analysis of the unscreened system we can address exciton propagation quantitatively.
Fig.~3~(c) shows normalized integrated intensity of the QW exciton PL 
in Sample 2, at $T=10$~K and for the same excitation powers in Figs. 3(a-b). 
Solid lines are a fit to a Gaussian distribution with half width at half maximum HWHM=$5.9$~$\mu$m, and a dashed line shows the laser intensity profile, characterized by a twice smaller  HWHM=$2.7$~$\mu$m.
It seems surprising, at first sight, that increasing the power does not help increasing the intensity of the light emitted by excitons away from the excitation spot.
To understand this effect, it is instructif to compare the corresponding spatial profiles of the emission energy (Fig. 3~(a-b)). 
The PL blue-shift is proportional to the local exciton density, and one can see that its spatial extension is much larger at $1$~mW that at $0.03$~mW. 
Indeed, we plot on top of the colormaps in Fig. 3(a) and (b), Lorentzian profiles with HWHM=$20$ and $14$~$\mu$m, which characterize the spatial distribution of the QW emission energy (even though the experimentally observed energy profiles have much more complex shapes).
We suggest, that the strong difference between the spatial profiles of energy and  intensity is a consequence of the strong density dependence of the exciton radiative lifetime. 
This dependence  is a prominent feature of IXs:  increasing exciton density  reduces exciton lifetime exponentially \cite{Lefebvre2004,Fedichkin2015}. 
It can be clearly seen in spatially integrated time-resolved experiments presented above, 
that as long as exciton emission exhibits a measurable blue-shift, 
the time-dependence of the emission intensity cannot be described by a single-exponential
decay  (see Fig. 4~(b-c)).
A quantitative study of this phenomenon combining spatially- and time-resolved $\mu$PL experiments would be of great interest, but it requires further experimental developments. 
Such studies are also mandatory to go further than our previous work \cite{Fedichkin2015} in the theoretical modelling of the exciton transport in this system, involving strong disorder, dipolar interactions and density-dependent exciton lifetimes.
Here we just stress that the exciton propagation length is determined by the spatial profile of the emission energy blue-shift, rather than by the spatial profile of the integrated emission intensity.

Let us now address  the temperature dependence of the exciton transport. 
In Fig.~2 we show on top of the colormaps Lorentzian profiles with HWHM values far above the excitation spot size.
These values are indicated for each measurement. The zero-density limit  (slightly different for the two samples as measured in time-resolved PL experiments) is shown by the dashed lines.
Here we neglect the combined effect of band-gap renormalisation and the exciton localisation on the zero-density emission energy. Strictly speaking, the variation of  $\pm 25$~meV can be expected \cite{Leroux1998PRB}.
This defines the precision of the propagation length determination, that we estimate to be of order of $\pm3$~$\mu$m.
In Sample 1, the signal was too low  at room temperature  to estimate the propagation length, while in Sample 2 the HWHM of the Lorentzian profile that defines the is propagation length is $\approx 12$~$\mu$m at room temperature. 
This is less than  maximum value measured at $10$~K (Fig. 3(a)), but is a remarkable result of this work.

Fig. 5(a) shows a set of spectrally-integrated and normalised spatial profiles 
of exciton emission in Sample 2
for five different temperatures from $10$ to $200$~K. 
A constant vertical offset is introduced for clarity,
and a dashed line shows the excitation spot profile, characterised by a HWHM= $2.7$~$\mu$m.
By comparing the measured intensity profiles (symbols) with a Gaussian profile of HWHM=$5.3$~$\mu$m  and with the laser spot profile, we conclude that the exciton emission intensity  pattern  is twice 
more extended than the laser spot and that the temperature dependence of this extension is very weak.
%
 This means that, with increasing temperature  up to $200$~K,
excitons are not lost via nonradiative recombination, but continue to contribute to the PL.
This is in contrast with Sample 1, where the non-radiative emission rate increases with temperature.
To quantify this effect we compare in Fig.~5(b) the total emission intensity measured in experiments like those shown in Fig. 2, but integrated both in energy and in space over a surface of $8000$~$\mu$m$^2$, for the two samples. 
One can see that, at highest excitation power, in Sample 2, the temperature dependence is indeed very weak, while in Sample 1 the intensity decreases by almost one  order of magnitude when increasing temperature by a factor of ten. 
This means that the suppression of the nonradiative losses is crucial for exciton propagation at room temperature.
At low excitation power the total integrated PL intensity is similar for the two samples. 
This suggests that even in Sample 2 nonradiative centers are not fully saturated in this weak excitation regime. The activation of nonradiative losses at low powers is a well-known effect in GaN and InGaN heterostructures \cite{Watanabe,Mickevicius}.

Despite low nonradiative losses, and their weak temperature dependence, 
exciton emission maps for Sample 2 at $30$ and $300$~K are very different (Fig.~2(d, f)).
The reason for this is the huge broadening of the exciton line.
This effect is illustrated in Fig. 5(c), where normalised exciton PL spectra measured $10$~$\mu$m away from the  laser spot are shown for three different temperatures.  
The peak emission energy is slightly different for the three spectra.
This  difference is related to the variation in the exciton density  but may  be contributed by the renormalization of the bandgap and the localisation effects at low temperature \cite{Varshni,KasiViswanath,Leroux1998PRB}.
The width of the spectrum increases with temperature. At room temperature the spectrum is so broad that
even  the first phonon replica cannot be distinguished.
At low temperatures and low densities (far away from the excitation spot) the linewidth is $w_{10K}\approx 40$~meV, limited by the in-plane disorder.  
Indeed, we estimated that monolayer fluctuations of the QW width in the presence of an electric field of $1$~MV/cm induce
 fluctuations of the in-plane potential $U_{rand}\approx 26$~meV,  and the fluctuations of the aluminium content in the barriers induce  $U_{rand}\approx 20$~meV per aluminium percent.
Under the same conditions at room temperature the line is so broad that  it is quite difficult to estimate the linewidth precisely: $w_{300K}\approx 130\pm 20$~meV.
The increase of the exciton linewidth with temperature has already been observed in similar structures \cite{Rossbach2014PRB}.
%
%
 For the highest  excitation power used in this work ($P=1.3$~mW) at $T=10$~K the linewidth reaches $80$~meV.
 %
 The exciton emission profile remains Gaussian as far as the exciton emission energy remains below $E=3.4$~eV, corresponding to a blue-shift $E_{BS}=400$~meV. 
 At higher carrier densities (only reached in the small area around the excitation spot at $P=1.3$~mW (see Fig. 2(d)), the line shape changes and its width increases dramatically, as expected at Mott transition \cite{Rossbach2014PRB}. 
 We estimate the corresponding Mott  density from the emission blue shift  $E_{BS}=400$~meV as
 $n_{M}=4 \times 10^{12}$~cm$^{-2}$, close to the values reported in \cite{Rossbach2014PRB}.
%
%
%

\emph{Conclusions.} 
We have demonstrated exciton transport in a GaN/(Al,Ga)N  QW grown on a GaN substrate.  
Propagation over $12$~$\mu$m at room temperature  and $20$~$\mu$m  at $10$~K is reported.
From the comparison of two nominally identical samples grown on GaN and sapphire, respectively,
we conclude that suppression of nonradiative losses at room temperature is the key to achieve this result. 
We have shown that the spatial distribution of excitons propagating away from the excitation spot is much wider than the corresponding distribution of the emission intensity,
due to the strong density dependence of the exciton emission rate.
The comparison of spatial and time-resolved experiments provides a clear evidence, that exciton transport is 
accompanied by photon propagation along the structure plane. This effect is much stronger 
in sapphire-grown samples but nevertheless cannot be neglected in GaN-grown samples.
It is responsible for the partial screening of the built-in electric field, so that the fully unscreened regime is
hard to reach in cw experiments.
 Our results  provide a rich playground for further understanding of exciton transport 
 and elaboration of new transport models, in order to optimize exciton propagation.

\emph{Acknowledgments} 
We are grateful to D. Scalbert and L. V. Butov for enlightening discussions.
This work was supported by  the European Union (EU ITN INDEX PITN-GA-2011-289968) and 
French National Research Agency (ANR OBELIX).

\bibliography{biblio}

\pagebreak

\section*{Figure captions} 
Figure 1: Sketch of the band structure with typical electron and hole wavefunctions  in the absence of photexcitation (a)
and under strong pumping (b). $E_0$ ($E_X$ ) is the exciton recombination energy in the zero (high) density limit; 
$E_{BS}$ is the population-induced blue shift $E_{BS} = E_X-E_0$.
Schematic representation of guiding effect in the structure grown on sapphire substrate (c) 
and reflection from GaN/air interface in the  structure grown on GaN substrate (d).

Figure 2: Color maps of the PL intensity taken at T =$30$, $80$ and $300$~K. Excitation power 
$P=2.3$~MW/cm$^2$.
The abscissa $r$ corresponds to the distance from the excitation spot. The ordinate corresponds to the energy spectrum. Dashed lines indicate the empty QW energy $E_0$ for each sample. Solid lines are Lorentzian profiles characterised by HWHM indicated on top of the images.

Figure 3: Power dependence of exciton transport in Sample 2 (GaN substrate). 
(a, b) Color maps of the PL intensity measured at $10$~K for two different excitation powers. Color scale is the same as in Fig. 2. Dashed lines indicate the empty QW energy $E_0$, 
as obtained from time-resolved experiments. 
Solid lines are Lorentzian profiles with HWHM of $20$ (a) and $14$~$\mu$m (b).
(c) PL intensity (symbols) corresponding to the images (a) and (b), integrated over energy from $2.85$ to $3.47$~eV and plotted as a function of the distance from the excitation spot.
Solid line is a Gaussian profile with $5.9$~$\mu$m variance, dashed line is the laser spot intensity profile.

Figure 4: Time-resolved exciton dynamics in Sample 2 (GaN substrate). 
(a) PL spectra measured at different time delays after the excitation pulse. 
(b) Exciton PL (symbols) integrated in energy 
from $2.8$ to $3.1$~eV as a function of the time delay after the laser pulse. Solid line is $10$~$\mu$s exponential decay.
(c) Exciton energy shift (symbols)  with respect to the lowest energy ($E_0=3$~eV), as a function of the time delay after the laser pulse. Solid line is $2$~$\mu$s exponential decay.

Figure 5: Temperature dependence of the exciton transport. 
(a) Intensity of  PL  (symbols)  in Sample 2 (GaN substrate), integrated from $2.85$ to $3.47$~eV as a function of distance from the excitation spot, at $5$ different temperatures. Solid lines are identical Gaussians with HWHM=$5.3$~$\mu$m.  Excitation power 
$P=2.3$~MW/cm$^2$.
(b) PL intensity integrated over energy (same as in (a)) and over surface (from $0$ to $100$~$\mu$m from the spot), as a function of temperature, for two different excitation powers ($15$~mW (red, black) and $0.5$~mW (cyan, magenta) for the two samples.
(c) Normalised PL spectra measured at three different temperatures at $10$~$\mu$m distance from the excitation spot.

\pagebreak
\section*{Figures} 

\begin{figure*}[ht!]
\includegraphics[clip,width=0.8\columnwidth]{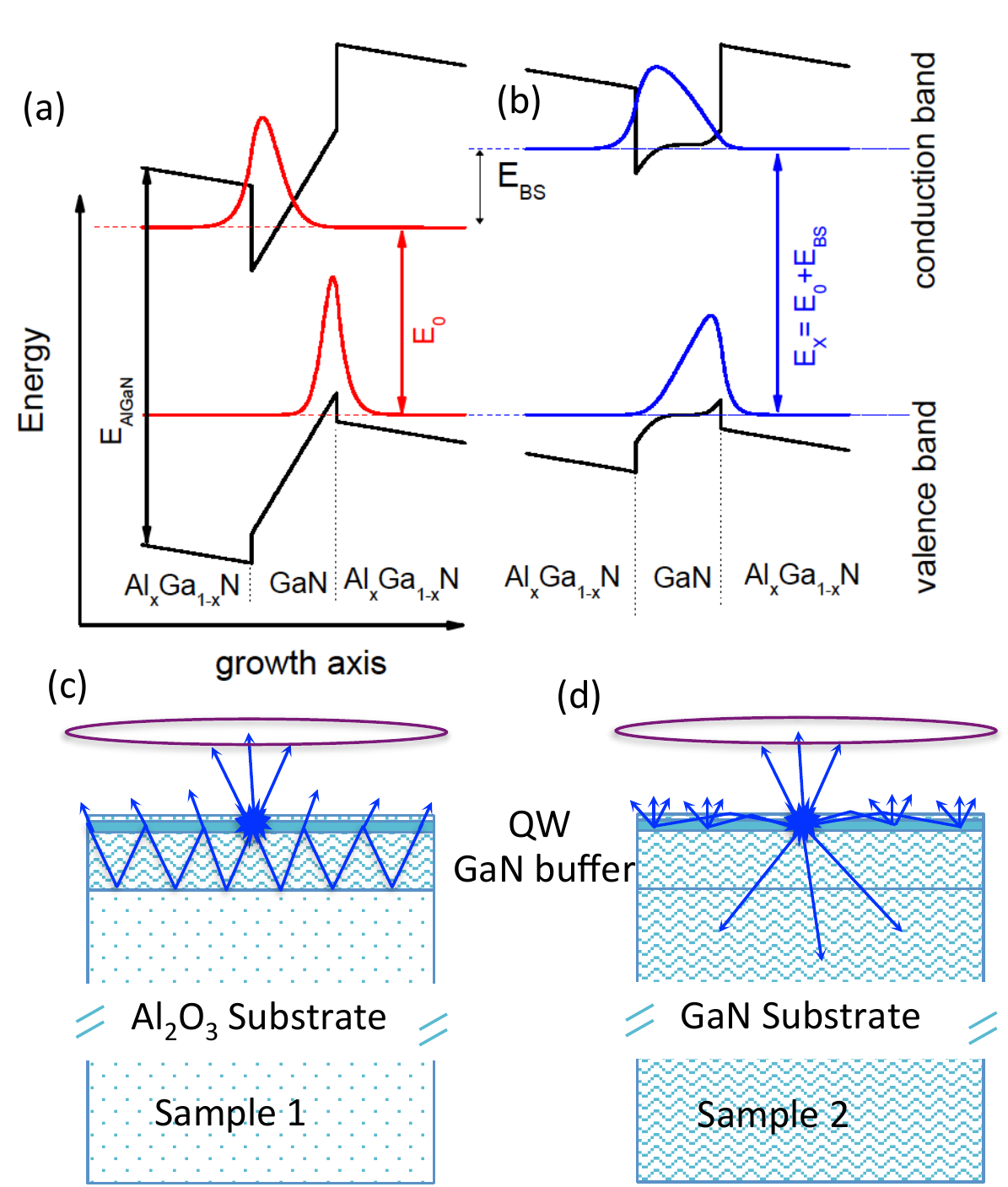} \label{fig:scheme} 
\caption{ }
\end{figure*}
\begin{figure*}[ht!]
\includegraphics[clip,width=1.0\columnwidth]{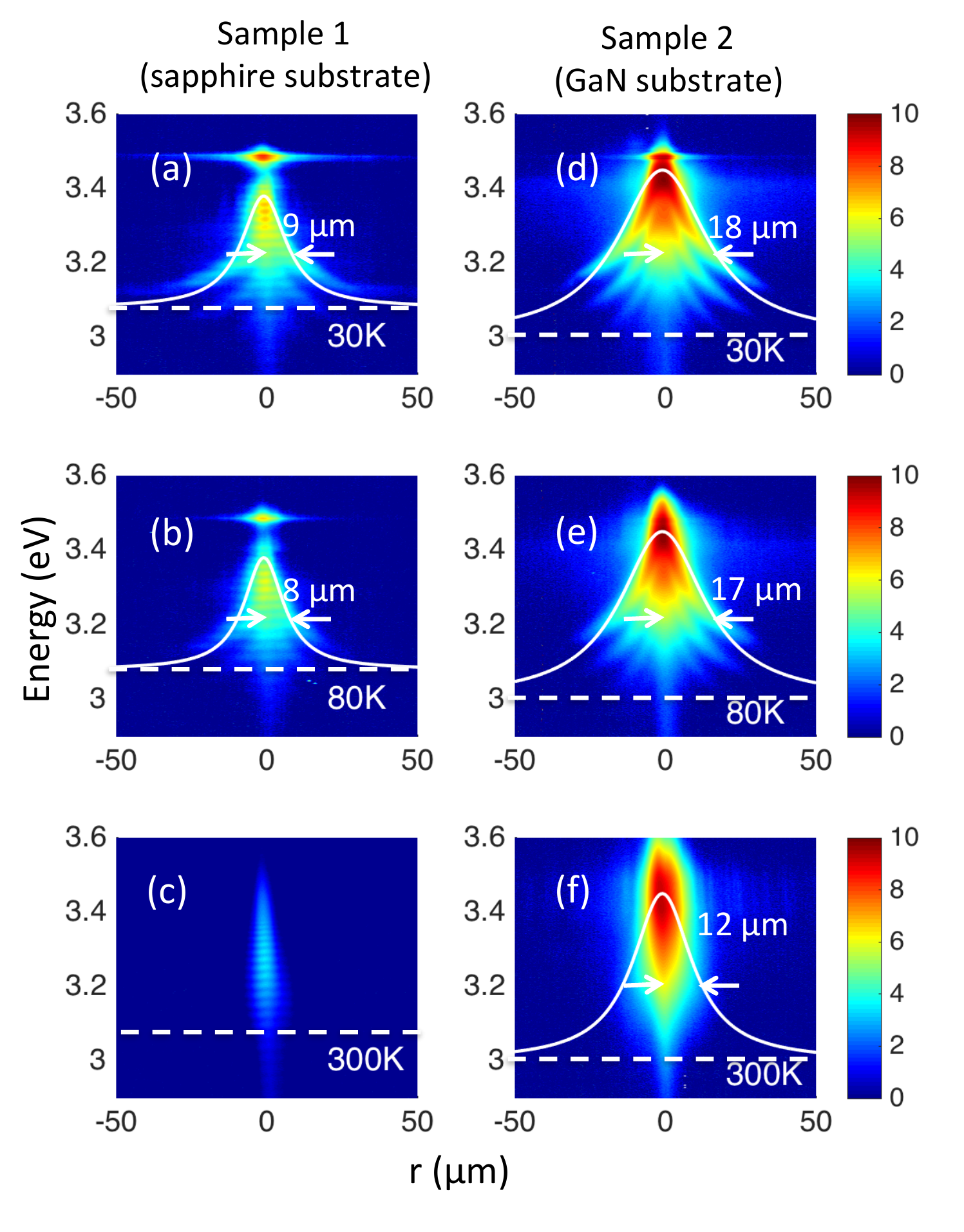} \label{fig:map} 
\caption{ }
\end{figure*}
\begin{figure*}[ht!]
\includegraphics[clip,width=0.8\columnwidth]{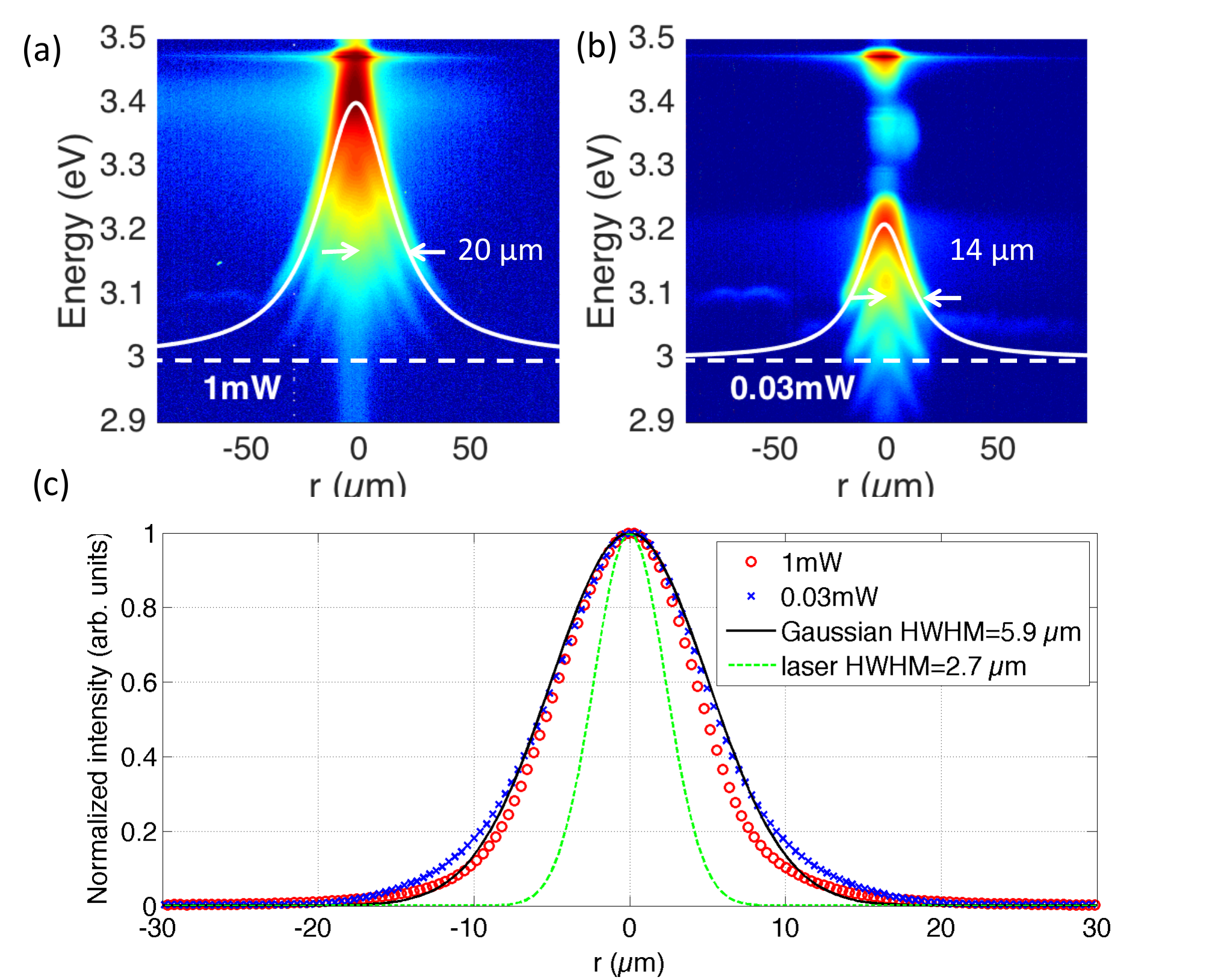} \label{fig:Pdep} 
\caption{ }
\end{figure*}
\begin{figure*}[ht!]
\includegraphics[clip,width=0.8\columnwidth]{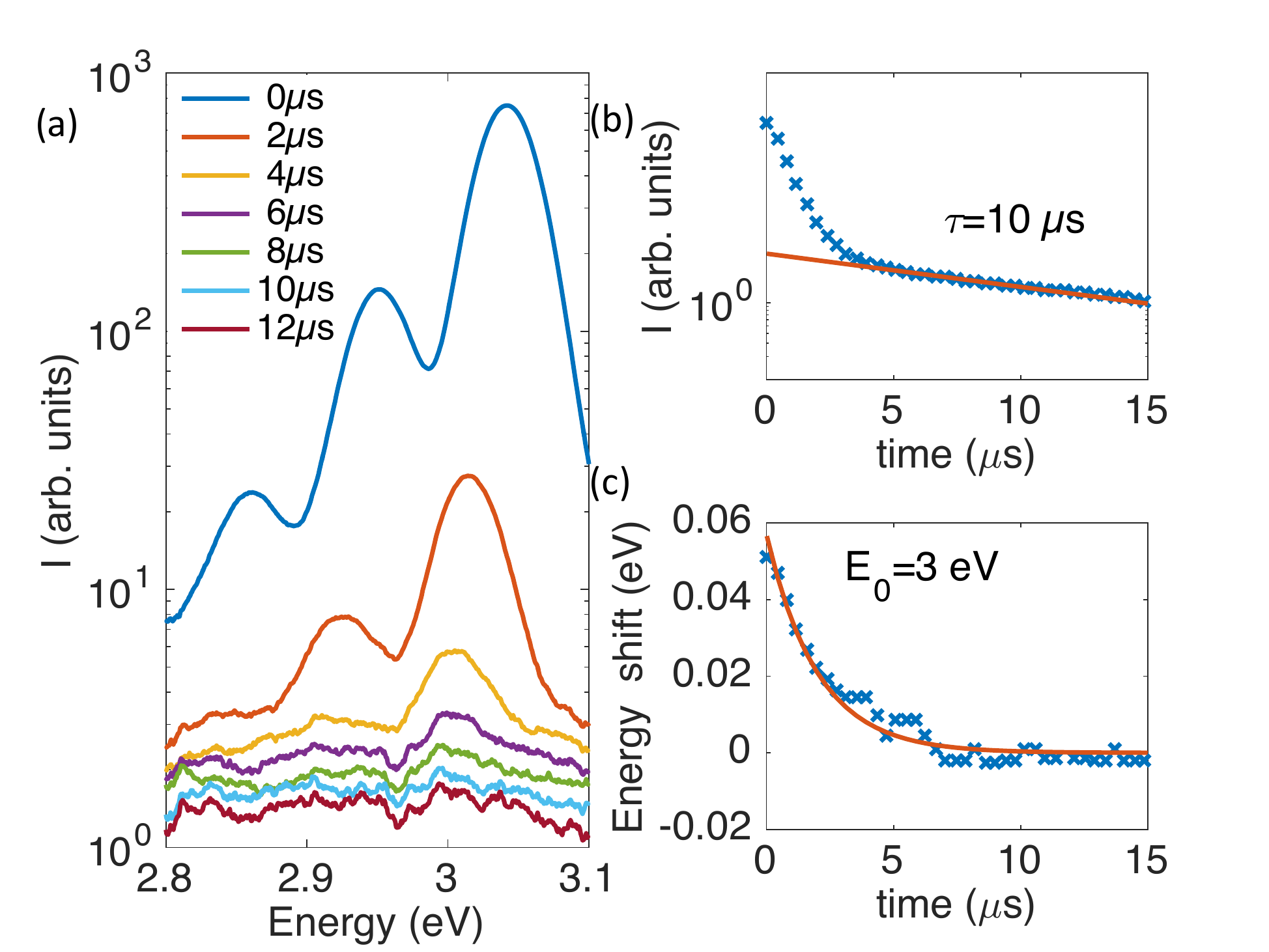} \label{fig:tr} 
\caption{ }
\end{figure*}
\begin{figure*}[ht!]
\includegraphics[clip,width=1.0\columnwidth]{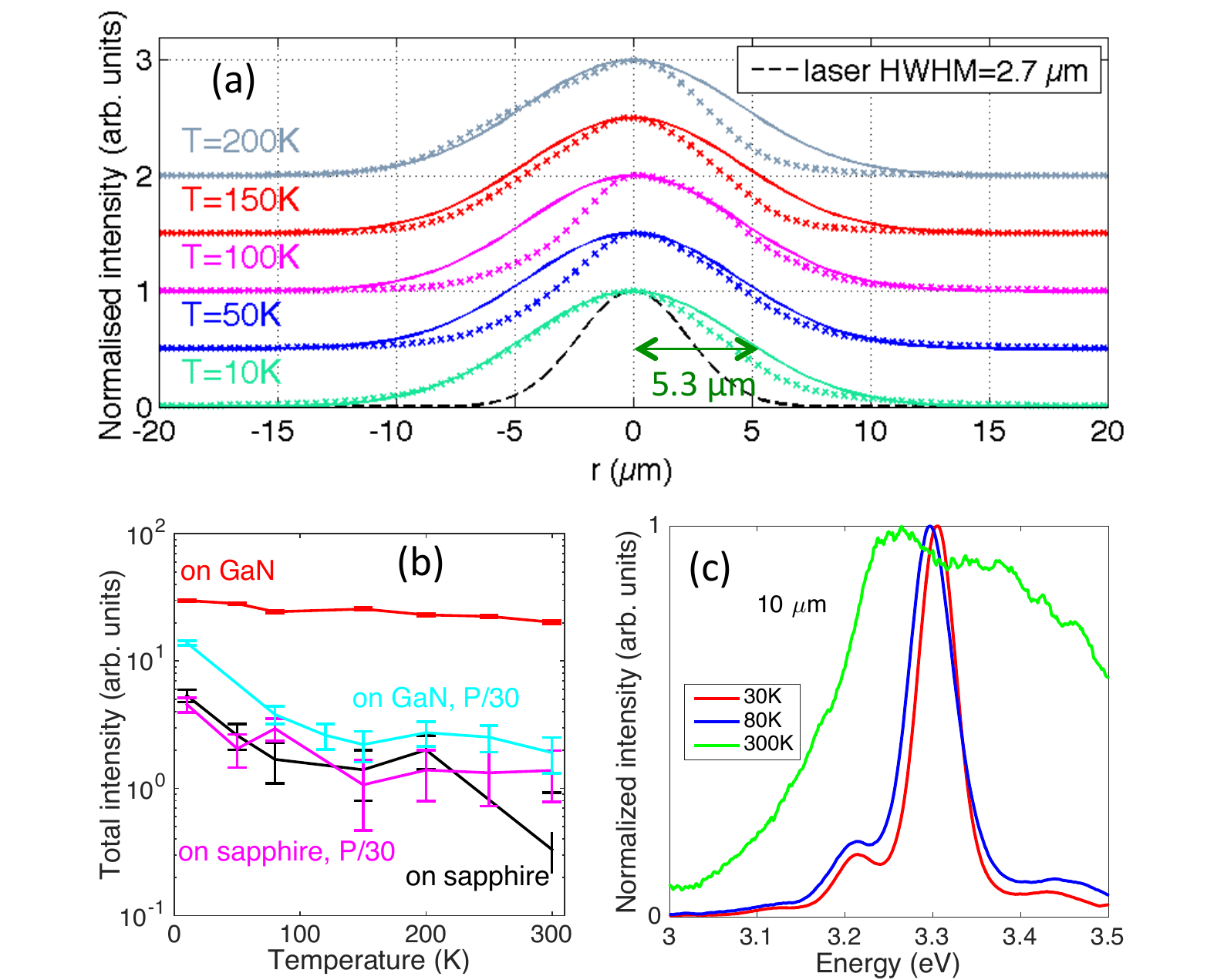} \label{fig:Tdep} 
\caption{ }
\end{figure*}

\end{document}